# Using Self-Organizing Maps for Sentiment Analysis

Anuj Sharma
Indian Institute of Management
Indore – 453331, INDIA
Email: f09anujs@iimidr.ac.in

Shubhamoy Dey
Indian Institute of Management
Indore – 453331, INDIA
Email: Shubhamoy@iimidr.ac.in

*Abstract*— Web 2.0 services have enabled people to express their opinions, experience and feelings in the form of user-generated content. Sentiment analysis or opinion mining involves identifying, classifying and aggregating opinions as per their positive or negative polarity. This paper investigates the efficacy of different implementations of Self-Organizing Maps (SOM) for sentiment based visualization and classification of online reviews. Specifically, this paper implements the SOM algorithm for both supervised and unsupervised learning from text documents. The unsupervised SOM algorithm is implemented for sentiment based visualization and classification tasks. For supervised sentiment analysis, a competitive learning algorithm known as Learning Vector Quantization is used. Both algorithms are also compared with their respective multi-pass implementations where a quick rough ordering pass is followed by a fine tuning pass. The experimental results on the online movie review data set show that SOMs are well suited for sentiment based classification and sentiment polarity visualization.

## INTRODUCTION

With the ever-growing popularity of Web 2.0 platforms and services, the Internet has emerged as a common medium to write reviews and share opinions for any product consumed or service received by users. Sentiment analysis, also known as opinion mining, attempts to address the problem of classifying opinionated texts (e.g. product reviews) according to their opinion polarity (positive or negative) (Pang et al, 2008). The sentiment based classification of online opinionated texts imposes several challenges which distinguishes it from topic based text classification. The variability and complexity of sentiment expressions, overlapping vocabularies, feature selection and domain dependency of sentiment features are the key issues in sentiment analysis (He et al, 2011; Bai, 2011).

This paper investigates the efficacy of Self-Organizing Maps (SOM) for sentiment based visualization and classification of online product reviews. SOM can transform high-dimensional data into two-dimensional representation and can enable automatic clustering of text documents, while preserving higher order topology. Specifically, this paper implements the unsupervised SOM algorithm to support sentiment based visualization and classification tasks. For supervised sentiment analysis, a competitive learning algorithm known as Learning Vector Quantization (LVQ) is used in this work. Both algorithms are also compared with their respective multi-pass implementations. The movie reviews dataset (Pang and Lee 2004) is used to evaluate performance of these different SOM implementations.

The rest of this paper is organized as follows: next Section presents related work on sentiment analysis, the methodology based on supervised and unsupervised SOM is described in the Section named methodology, the experimental result Section discusses our results, and the last Section concludes the paper.

**RELATED WORK**

There have been a considerable number of studies on sentiment analysis and opinion mining, most of which utilize machine learning, dictionary, statistical, and semantic based approaches. Supervised machine learning classifiers like Naïve Bayes, Support Vector Machine, Maximum Entropy, Decision Tree, Centroid classifier, K-nearest neighbor, Neural Networks, winnow classifier etc., are quite popular for sentiment based classification of online reviews (Sharma et al, 2012a, 2012b; Pang et al, 2002, 2004; Tan et al, 2008). Dictionaries of subjective words with some weighting schemes are also proposed in some studies to quantify polarities of sentiment expressing words in texts for sentiment analysis (Amati et al, 2008). Statistical approached apply statistics like Pointwise Mutual Information (PMI) based on co-occurrence of words to derive sentiment polarity of features (Turney, 2002). The Semantic approach exploits various semantic relationships between sentiment words like synonyms and antonyms, to calculate sentiment polarities for mining and summarizing customer reviews (Hu, 2004).

The machine learning approaches that belong to supervised text classification techniques have shown good accuracy. But, the performance of machine learning approaches is heavily dependent on the selected features, the quality and quantity of training data and the domain of the data. The additional learning time required by the machine learning techniques is also a prominent issue since lexicon or semantic approaches take less time.

The Naïve Bayes classifier is a probabilistic classifier based on Bayes' theorem with strong feature independence assumptions. In real life, the sentiment features are seldom independent (Pang, 2002), severely limiting its applicability. The complexity of SVM due to its convoluted training and categorizing algorithms is a major drawback. The over-fitting issues with small training sets also limits the use of SVMs in many real world applications. The k-nearest neighbor (k-NN) classifier is easy to implement but becomes computationally intensive with growing size of the training set (Han et al, 1999). The decision tree and rule induction classifiers are simple to understand and explain but their performance suffer when the number of distinguishing features is large (Quinlan, 1993).

To address the above issues, this paper implements both supervised and unsupervised SOM based approaches for sentiment based classification and compares their performances in the same domain. The findings will help gaining insight into the strengths and limitations of different SOM implementations for sentiment analysis.

**METHODOLOGY**

*Data Pre-processing*

The data pre-processing involves collecting opinionated text documents, performing initial data cleaning and transforming relevant text data into the input variables form as needed by self-organizing maps. In this study we have performed tokenization, stemming and stop word removal in data pre-processing. We have used Vector Space Model (VSM) to generate the bag of words representation for each document. In VSM, the weights represent the *tf·idf* score which is computed from term frequency (*tf*) and inverse document frequency (*idf*). The *tf·idf* weight for term *j* in document $D_i$ is computed as

$$t_{ij} = tf_{ij} \times idf_j = \log(1 + f_{ij}) \times \log(\frac{n}{df_j}) \qquad (1)$$

Where $f_{ij}$ is frequency of term *j* in document *I* and $df_j$ is number of document having term *j*.

*Information Gain Feature Selection*

Information gain (IG) as a feature goodness criterion has shown consistent performance for sentiment-feature selection (Tan et al, 2008). A feature from large feature space is selected based on its impact on decreasing the overall entropy. Entropy is the expected information needed to classify an instance document. The attributes ranked as per high to low IG score will minimize the overall information necessary to classify instances into predefined classes. The information gain of a feature *j* over all classes is given by:

$$IG(j) = -\sum_{i=1}^{|C|} P(c_i) \log_2 P(c_i) + P(j)\sum_{i=1}^{|C|} P(c_i \mid j) \log_2 P(c_i \mid j) + P(\bar{j})\sum_{i=1}^{|C|} P(c_i \mid \bar{j}) \log_2 P(c_i \mid \bar{j}) \qquad (2)$$

Where, $P(c_i)$ is the probability that a random instance document belongs to class $c_i$. $P(j)$ is the probability of the occurrence of the feature *j* in a randomly selected document. $P(c_i|j)$ is the probability that a randomly selected document belongs to class $c_i$ if document has feature *j*. $P(\bar{j})$ is the probability of that *j* does not appear in a random document. $P(c_i \mid \bar{j})$ is the probability that a random document belongs to class *c* if *j* does not occur in it. The top ranked features are selected based upon information gain feature selection in this study.

*Self-Organizing Maps (SOM)*

The self-organizing map (SOM) is an unsupervised competitive ANN based on the winner-takes-all (WTA) principle (Kohonen, 1995). A typical SOM is made up of a vector of nodes for input (input neurons layer), an array of nodes as output map, and a matrix of connections between each output layer unit and all the input layer units. The input nodes receive vectorized input data patterns and propagate them to a set of output nodes, which are arranged according to map topology (generally two-dimensional grid) forming so called Kohonen's layer. Kohonen's principle of topographic map formation, determines how the spatial location of an output node in the topographic map corresponds to a particular feature of the input data pattern (Merkl, 1998).

Each of the output nodes $j$ is associated with a weight vector $w_j$ of the same dimension as the input data vectors and a position in the map space. The popular arrangement of output nodes is a regular spacing in a hexagonal or rectangular grid. The procedure for placing a vector from input data space onto the map is to first find the node with the spatially closest weight vector to the input data space vector. Once the closest node is located it is assigned the values from the vector taken from the data space. For Sentiment based classification, each vector represents an online review document, while the output node represents the sentiment category that the review is classified to.

The SOM learns through self organization of random nodes whose weights are associated to the layers of neurons. These weights are updated at every epoch during the training process. The change in weights depends upon the similarity or spatial closeness between the input data pattern and the map pattern (Michalski et al, 1999).

Let $x_i \in \mathbb{R}^M$, $1 \leq i \leq N$, be the randomly chosen training feature vector representing document $i$ in the corpus, where $M$ is the number of indexed features and $N$ is the number of documents in input dataset. These vectors are used to train the map which consists of a regular spacing grid of processing units called neurons. Each neuron in the map has $M$ connections (synapses). The synaptic weight vector $w_j$ is associated with the $j$th neuron in the map having $J$ neurons where, $w_j = \{w_{jm}/1 \leq m \leq M\}$, $1 \leq j \leq J$. The SOM learning algorithm used in this paper includes the following steps:

Step 1. Initialize weights of neurons in map.

Step 2. Randomly select a training vector $x_i$ from the corpus.

Step 3. Determine the neuron *c*, having the highest activity level with respect to $x_i$.

The activity level of a neuron is represented by the Euclidean distance between the input pattern and that neuron's weight vector. The neuron having the highest activity level is referred as best matching unit (BMU). Hence, the selection of BMU may be written as:

$$c : \| x_i(t) - w_c(t) \| = \min_j (\| x_i(t) - w_j(t) \|) \tag{3}$$

The operator $\| \cdot \|$ denotes Euclidean vector norm as the neuron *j* with synaptic weights $w_j$ is closest to $x_i$. The discrete time notation *t* denotes the current training iteration.

Step 4. Update the synaptic weights of the BMU and its neighbors in order to move the BMU closer to the input vector.

The map adapts to the input pattern in all iterations of learning, and adaption is performed as a gradual reduction of the spatial distance between the input vector and BMU's weight vector. The learning rate parameter *α(t)* controls the amount of adaptation that gradually decreases during the course of training.

A neighborhood kernel function is used to determine the spatial range of neurons around the BMU that are subject to adaptation. Based upon the spatial distance between BMU and neighboring neurons, the neighborhood function determines the width of the adaptation kernel. This ensures that neurons spatially close to the BMU are adapted more strongly than neurons further away. A Gaussian function (Eq. (4)) can serve as the neighborhood function with $r_i$ representing the two-dimensional vector pointing to the location of neuron *j* within the map, and $\| r_j - r_c \|$ denoting the distance between neuron *j* and BMU (i.e., *c*) in terms of the output space.

$$h_{c(x_i), j} = e^{-\| r_j - r_c \|^2 / 2\sigma^2(t)} \tag{4}$$

The time-varying parameter σ guides spatial width of adaptation in such a way that a wide area of the output space is subject to adaptation at the beginning of training. The number of neighboring neurons affected by adaptation is reduced gradually during the training process. The learning rule for any output neuron *j* can be represented as

$$w_j(t+1) = w_j(t) + \alpha(t) h_{c(x_i), j}(t) \cdot \left[ x_i(t) - w_j(t) \right] \tag{5}$$

Step 5. Increase time stamp *t* representing training iteration. If t reaches the preset maximum training time *T*, stop the training process; otherwise decrease *α*(*t*) (0< *α*(*t*) <1), and the neighborhood size, and go to Step 2.

The current study has implemented SOM algorithm as a baseline model and its multi-pass implementation for sentiment analysis. The multi-pass SOM is recommended for better results where two passes are executed on the same underlying model. The first pass is executed as a rough ordering pass with large neighborhood, learning rate and small training time. The second pass is performed as the fine tuning pass that has a longer training time, small initial neighborhood and smaller initial learning rate (Kohonen et al, 1996).

SOM have two advantages over other clustering methods: non-linear projection of the input space and cluster topology preservation. Thus, SOM can extract inherent non-linear relationships amongst documents, and similar clusters of documents are mapped close to each other revealing higher level (i.e. emergent structures) (Ultsch et al, 2005). In Emergent Self Organizing Maps (ESOM) the cluster boundaries are 'indistinct', the degree of separation between 'regions' of the map (i.e. clusters) being depicted by 'gradients' (Ultsch et al, 2005). These attractive features of SOM have prompted researchers to use SOM for text clustering and visualization (Yen et al, 2008; Feng et al, 2010; Dey, 2010).

**Learning Vector Quantization (LVQ)**

The Learning Vector Quantization (LVQ) algorithm is the supervised version of the Kohonen's SOM model designed for pattern classification tasks (Kohonen, 1995). The application of the LVQ algorithm to sentiment classification of user generated text has not been explored so far.

The LVQ algorithm is based on neural competitive learning, which enables defining a group of class labels on the input data space using reinforced learning. LVQ also transforms high dimensional input data into a two-dimensional map like SOM, but without taking into consideration the topology of input data. For this transformation, LVQ utilizes pre assigned class labels to documents, thus minimizing the average expected misclassification probability. Hence, unlike the SOM, where clusters are generated by unsupervised manner based on feature-vector similarities, the LVQ categories are predefined.

LVQ is a single-layer network where, the number of output neurons is equal to number of predefined classes (e.g., *J*=2 for binary sentiment classification). The weight vectors associated with each output neuron are called codebook vectors. Each class of input space is represented by its own set of codebook vectors and one codebook vector per class is used for classification.

Any output neuron $j$, whose weight vector matches the input pattern $x_i$ as per a spatial distance definition is the winning unit (BMU), and its weights are updated during training (Goren-Bar et al, 2005).

The LVQ algorithm is a competitive one, as for each input training vector output neurons compete among themselves in order to find the BMU as per Euclidean distance. The LVQ learning rule modifies weights of only the BMU, as it has the smallest Euclidean distance with regard to the input vector. The LVQ algorithm implemented in this paper includes the following steps:

Step 1. Initialize the codebook vectors $w_i$.

Step 2. Randomly select a training vector $x_i$ from the corpus.

Step 3. Determine the BMU (winner unit) closest to the input vector. The codebook vector $w_c$ associated with BMU has the smallest Euclidean distance with regard to the input vector $x_i$.

$$c : \| x_i(t) - w_c(t) \| = \min_j (\| x_i(t) - w_j(t) \|) \qquad (6)$$

Step 4. Update the synaptic weights of the BMU.

- For correction classification (i.e. if codebook vector $w_c$ and input vector $x_i$ belong to the same class)

$$w_j(t+1) = w_j(t) + \alpha(t) \left[ x_i(t) - w_j(t) \right] \qquad (7)$$

- For incorrect classification (i.e. if codebook vector $w_c$ and input vector $x_i$ belong to different class)

$$w_j(t+1) = w_j(t) - \alpha(t) \left[ x_i(t) - w_j(t) \right] \qquad (8)$$

Step 5. Increase time stamp $t$ representing training iteration and reduce the learning rate $\alpha$. Repeat from step 2 until the neural network is stabilized or until a fixed number of iterations have been carried out.

The current study has implemented LVQ1 with single BMU selection, Optimized LVQ1 (OLVQ1) with separate learning rate for each codebook vector and Multi-Pass LVQ. Multi-Pass LVQ implements two passes where, a quick rough pass is made on the model using OLVQ1, then a long fine tuning pass is made on the model with LVQ1 (Kohonen, 1990).

**EXPERIMENTAL EVALUATION**

*Datasets and Performance Evaluations*

To evaluate the performance of SOM and LVQ for sentiment classification, this paper experiments with a data set of classified movie reviews (Pang and Lee, 2004). The data set contains 1,000 positive and 1,000 negative reviews and is a benchmark dataset for sentiment analysis tasks. This work has adopted classification accuracy, precision, recall and $F$1 Score to evaluate the performance of sentiment classification.

*Experimental Results*

We attempted to test the effectiveness of ESOM in visualizing sentiment polarity in classified movie reviews data set. Input features were selected using Information Gain criteria, and experiments were conducted with 50 to 1000 selected features.

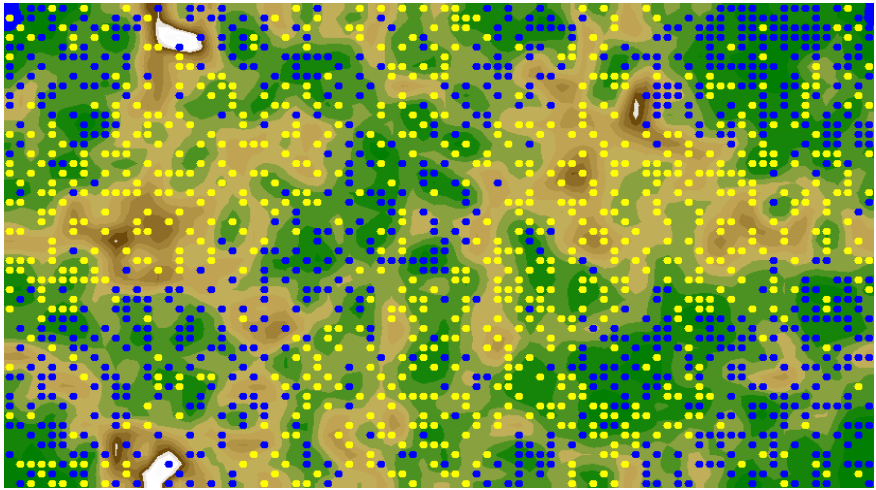

Fig. 1. ESOM of Vectors with 100 Features

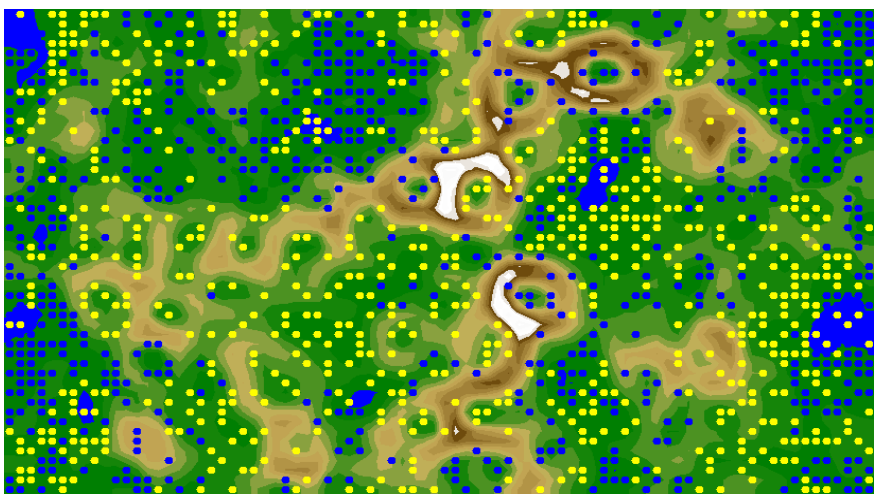

Fig, 2. ESOM of Vectors with 700 Features

The resulting visualization maps for 100 and 700 features are shown in Figures 1 and 2 respectively. In the maps, the *Blue* dots represent positive polarity and *Yellow* dots represent negative polarity reviews (as classed in the movie reviews dataset). It can be observed from Figures 1 and 2 that SOMs are able to distinguish between reviews with positive and negative sentiment in the form of clusters. It is also seen that more distinct and well separated clusters are formed when vectors with 700 features are used to represent the reviews.

Java based implementations on Microsoft Windows platform were used to implement all the versions of SOM and LVQ algorithms. The number of indexed terms $N$ to be used in input document vector was selected from 50 to 1000. All the algorithms were experimented with taking random training data, proportional initialization mode with linear decay learning function. For SOM, this paper has used Gaussian function as the neighborhood function with neighborhood size 8 for a hexagonal map topology as suggested in (Kohonen, 1995). The learning rate, $\alpha$ was selected 0.3 for SOM, LVQ1, OLVQ1 and for the first pass of multi-pass SOM and multi-pass LVQ. For second pass, the learning rate was kept as small as 0.05. The number of iteration was varied from 1000 to 10000 to obtain a training map. All experiments were validated using 10-fold cross validation.

Figures 3, 4, 5, and 6 shows comparison of both SOM and LVQ implementations on accuracy, precision, recall, and $F1$ score of sentiment based classification on movie reviews dataset. The Multi-Pass LVQ performed best among all algorithms with 89.1% overall accuracy. The (classic Kohonen's) SOM was the worst performer among all.

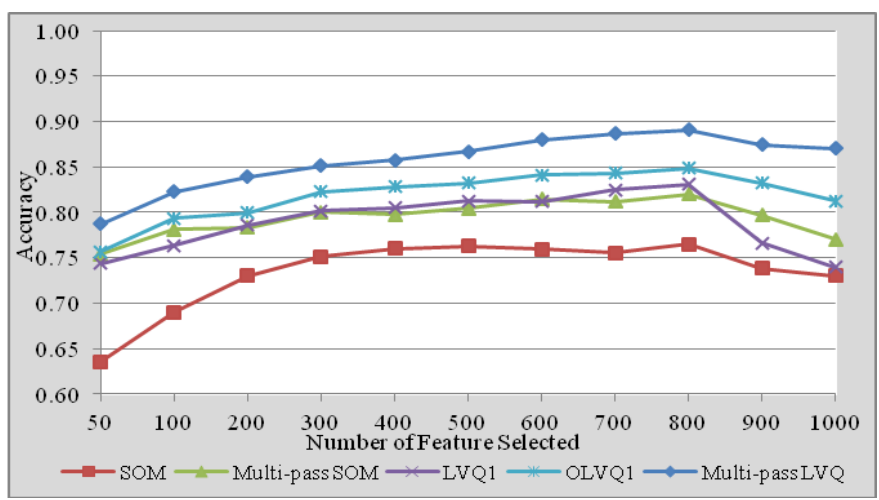

Fig. 3. Performance Comparison on Accuracy

Almost all implementations of SOM and LVQ the best performances were obtained for 700-800 selected features. These results are consistent with previous studies related to sentiment analysis, as a few sentiment bearing features can express enough polarity for the whole document to be classified (Turney, 2002). Degradation in the performance of sentiment based classification can be seen beyond 900 features as the feature matrix becomes sparse, affecting the organization of maps. The same algorithms with larger number of features (i.e. 2000 to 10000) were also executed in this study and it was observed that performance of the classifier degraded significantly due to high dimensionality of the input space.

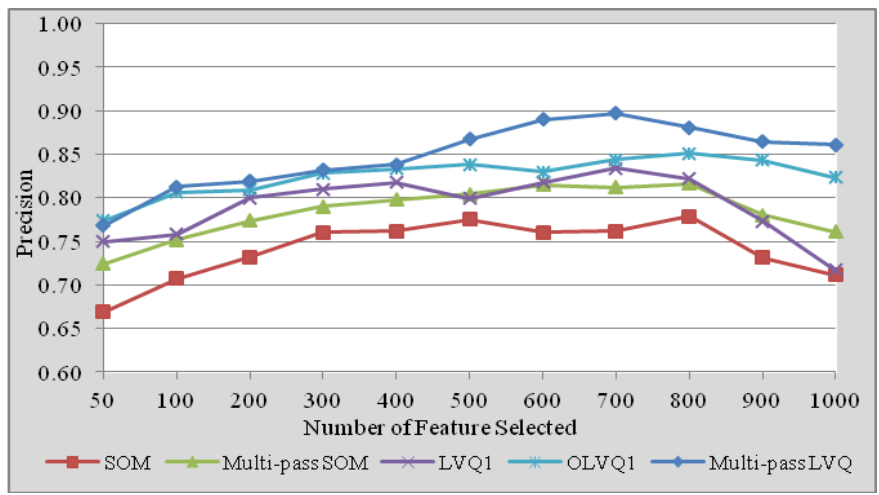

Fig. 4. Performance Comparison on Precision

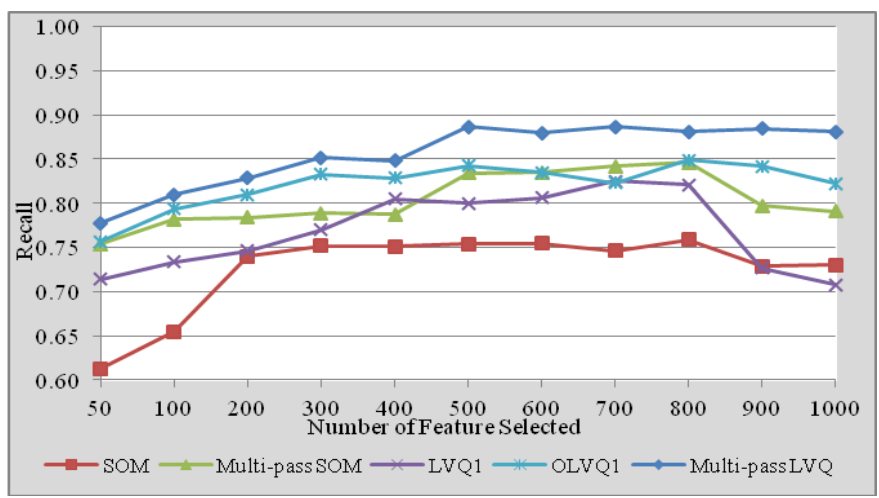

Fig. 5. Performance Comparison on Recall

As such, with respect to learning methods, LVQ and SOM are all suitable for sentiment analysis as their best performance results are comparable with other studies using other machine learning classifiers (Pang et al, 2004) and other unsupervised learning methods

(Turney, 2002). The time and computation resources sacrificed to execute the two passes in multi-pass implementations are justified as they have shown better performance as compared to their respective single-pass implementations. The OLVQ1 has worked well with 84.9% best overall accuracy, which has ranked it as the second best algorithm among all in this study. Though this study did not compare the execution time of all the algorithms but it is worth noting that OLVQ1 was the quickest among all in convergence.

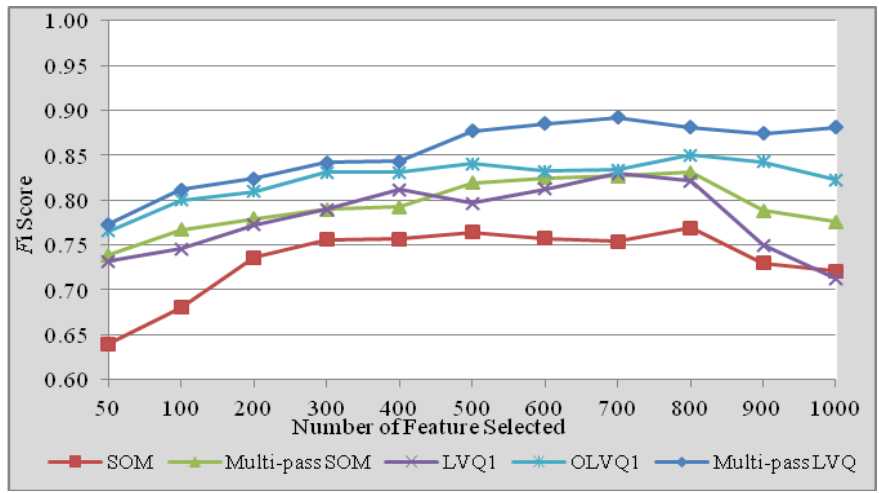

Fig. 6. Performance Comparison on $F$1 Score

**CONCLUSION**

This paper proposes a sentiment classification model using SOM and LVQ algorithms. Specifically, this paper investigates performance of different implementations of SOM for sentiment based visualization and classification of online product reviews. The supervised learning algorithms (LVQ, OLVQ1, and Multi-pass LVQ) have been found to perform better than unsupervised learning algorithm for sentiment analysis purpose. Experiments have been performed to support sentiment based visualization and classification tasks. Both SOM and LVQ algorithms have also been compared with multi-pass implementations, which performed better than their respective single-pass implementations. The experimental results on the online movie review data set show that SOMs are well suited for sentiment based classification and sentiment polarity visualization.